\documentclass{moriond}
\usepackage{amssymb}

\def\be{\begin{equation}}
\def\ee{\end{equation}}
\def\bea{\begin{eqnarray}}
\def\eea{\end{eqnarray}}

\usepackage{amsmath}
\usepackage{amsfonts}
\usepackage{amssymb}
\usepackage{latexsym}
\usepackage{graphicx}
\usepackage{float}
\usepackage{epsfig}
\usepackage{hyperref}
\usepackage[section]{placeins}

\newcommand{\bmat}{\left ( \begin{array}{cc} 
}
\newcommand{\emat}{\end{array} \right )}

\newcommand{\Tr}{\textrm{Tr}\,}

\newcommand{\eins}{\leavevmode\hbox{\small1\kern-3.8pt\normalsize1}}
\newcommand{\vect}{\left ( \begin{array}{c}}
\newcommand{\evect}{ \end{array} \right )}

\begin{document}
\vspace*{4cm}
\title{Discretization effects in $N_c=2$ QCD and Random Matrix Theory~\footnote{Talk presented at the 50th Rencontres de Moriond (QCD and High Energy Interaction Session), La Thuile, 24 March 2015.}}

%\author{ A.B. AUTHOR }
\author{
Mario Kieburg$^{1,2} $, Jacobus Verbaarschot $^2$, \\ \underline{ Savvas Zafeiropoulos}$^{2,3}$\,\footnote{Speaker.}\\
%
%\address{
$^1$~Fakult\"at f\"ur Physik, Universit\"at Bielefeld, Postfach 100131, 33501 Bielefeld, Germany \\
$^2$~Department of Physics and Astronomy,  Stony Brook University, NY 11794-3800, USA  \\
$^3$~ Institut f\"ur Theoretische Physik, Goethe-Universit\"at Frankfurt,\\
Max-von-Laue-Str.~1, 60438 Frankfurt am Main, Germany \\
}
%}
%\address{Department of Physics, Theoretical Physics, 1 Keble Road,\\
%Oxford OX1 3NP, England}
%\vspace*{1cm}
\maketitle\abstracts{
\vspace*{0.5cm}
We summarize the analytical solution of the Chiral Perturbation Theory for the Hermitian Wilson Dirac operator of $N_c=2$ QCD with quarks in the fundamental representation.
Results have been obtained for the quenched microscopic spectral density, the distribution of the chiralities over the real modes and the chiral condensate.
The analytical results are compared with results from a Monte Carlo simulation of the corresponding Random Matrix Theory. }

\section{Introduction}
The case of $SU(2)$ QCD with fundamental quarks has attracted a great deal of interest in the past decades since it is a very interesting theory per se but it also shares many of the salient properties of $SU(3)$ QCD, such as confinement and chiral symmetry breaking.
An important difference between $SU(2)$ and $SU(3)$ is that $SU(2)$ is pseudoreal, so that for an even number of flavors it is not hindered by the notorious sign problem at non zero chemical potential. Therefore, it is in an interesting playground for finite density studies, and one can study the phase diagram of $SU(2)$ QCD in the $T-\mu$ plane \cite{Kogut,Hands} and get some insight in the qualitative properties of the phase diagram of QCD.
One can get an understanding of the systematics but one needs to bear in mind that the two theories have quantitative differences. For $SU(2)$ QCD  the baryons are of bosonic nature and as a result this theory has a superfluid phase that is not encountered in ordinary QCD.
Moreover, for beyond the Standard Model studies dedicated to the search of the conformal window, it has been shown that one can enter the conformal window with considerable less flavors compared to the $SU(3)$ theory \cite{kari}.
All these non-perturbative studies have been mainly through lattice simulations employing one of the most prominent discretizations, that of Wilson fermions. This discretization explicitly breaks chiral symmetry and consequently the low lying part of the Dirac spectrum gets affected in a very particular way.

We study the discretization (cutoff) effects on the Dirac spectrum by employing the methods of Wilson Chiral Perturbation Theory (WchPT) and Wilson Random Matrix Theory (WRMT). This work extends previous work on $SU(3)$ QCD \cite{DSV10,ADSV,KVZPRL,lat12,KVZlong,KVZ-SU2} and we point out the similarities and the differences among the two theories.
\section{Wilson chPT}
Initially we concentrate on the study of the Hermitian Wilson Dirac operator $D_5=\gamma_5 D_W$ where $D_W$ is given by 
\be
D_W =\frac{1}{2}\gamma_\mu(\nabla_\mu+\nabla_\mu^*)
     -\frac{a}{2}\nabla_\mu\nabla_\mu^*.
\ee
We work in the $\epsilon-$ regime of WchPT where the mass $m$, the axial mass $z$ and the lattice spacing $a$ scale with the lattice volume such that the quantities 
\be
 mV\Sigma, \quad  zV\Sigma, \quad {\rm and} \quad a^2VW_8
\ee
are kept fixed ($\Sigma$ is the chiral condensate and $W_8$ is the leading LEC). In this regime the partition function of WchPT factorizes and its mass dependence is given by a unitary matrix integral since the fluctuations of the zero momentum pions dominate completely the ones with non-zero momentum. The chiral partition function for $N_f$ flavors reads \cite{SS}
\be
Z_{N_f}(m,z;a) =   \int_{SU(2N_f)/Sp(2N_f)} \hspace{-1mm} d U e^{\frac{m}{2}\Sigma V{\rm Tr}(U+U^\dagger)+
\frac{z}{2}\Sigma V{\Tr}(U-U^\dagger)
-a^2 V W_8{\Tr}(U^2+{U^\dagger}^2) }.
\ee

To access the spectral properties of the Dirac operator we employ the supersymmetric method. Adding an additional fermionic quark as well as an additional bosonic quark, with source terms in the mass as well as in the axial mass, one can differentiate with respect to these sources and compute the Green's function of this theory in the standard way. By setting the source terms to zero one recovers the original partition function.
In this proceeding we present our results graphically and we refer the reader to the original work for detailed derivations \cite{KVZ-SU2}.

The first spectral observable that we consider is the quenched resolvent of $D_5$ defined as
\be
G_5(m,z) = \frac 1V {\rm Tr} \frac 1{D_5+z}
\ee
whose discontinuity yields the spectral density of $D_5$
\be
\rho_5(\lambda) = \frac 1 \pi {\rm Im} G_5(\lambda+i\epsilon).
\ee
The spectral density of $D_5$ for $N_c=2$ exhibits a striking difference compared to $N_c=3$ QCD. We see that there is a non-uniform convergence to the continuum limit \cite{JV-su2} since there is a discontinuous jump by a factor of $1/\sqrt{2}$ which is persistent for arbitrarily small values of the lattice spacing as it is clearly shown in Fig.~1. 

At the moment we have not been able to derive analytical results for the real eigenvalue density of $D_W$ but we have obtained stringent upper and lower bounds of this distribution.
The first distribution has been coined as the distribution of chiralities over the real modes of $D_W$ and is defined as
\be
\rho_\chi(\lambda) =
\sum_{\lambda_W \in {\mathbb R} }
\delta( \lambda +\lambda_W)
{\rm sign }[
\langle \lambda_W | \gamma_5 | \lambda_W \rangle],
\ee
and together with 
\be
\rho_5(\lambda_5=0, m;a)= \left \langle \sum_k \delta(\lambda_k^5(m)) \right \rangle=
\left \langle \frac{\delta(\lambda_k^W+m )}{|\langle k|\gamma_5|k\rangle| }
\right \rangle,
\ee
we have the following inequality for the density $\rho_{\rm real}(\lambda)$ of the real modes of $D_W$, 
\be 
\rho_\chi(\lambda) \le \rho_{\rm real}(\lambda) \le \rho_5(\lambda_5=0, m=\lambda).
\label{inequality}
\ee
In Fig.~\ref{bounds} we see that for large values of the lattice spacing these distributions are bounding 
the real eigenvalue density from above and below. While for values of $a$ close to zero we see that $\rho_{\chi}$ is a very good approximation of $\rho_{\rm real}$ .

\begin{figure}
 \label{rho5}
\centering
\begin{minipage}{.5\textwidth}
  \centering
  
  \includegraphics[width=.65\linewidth]{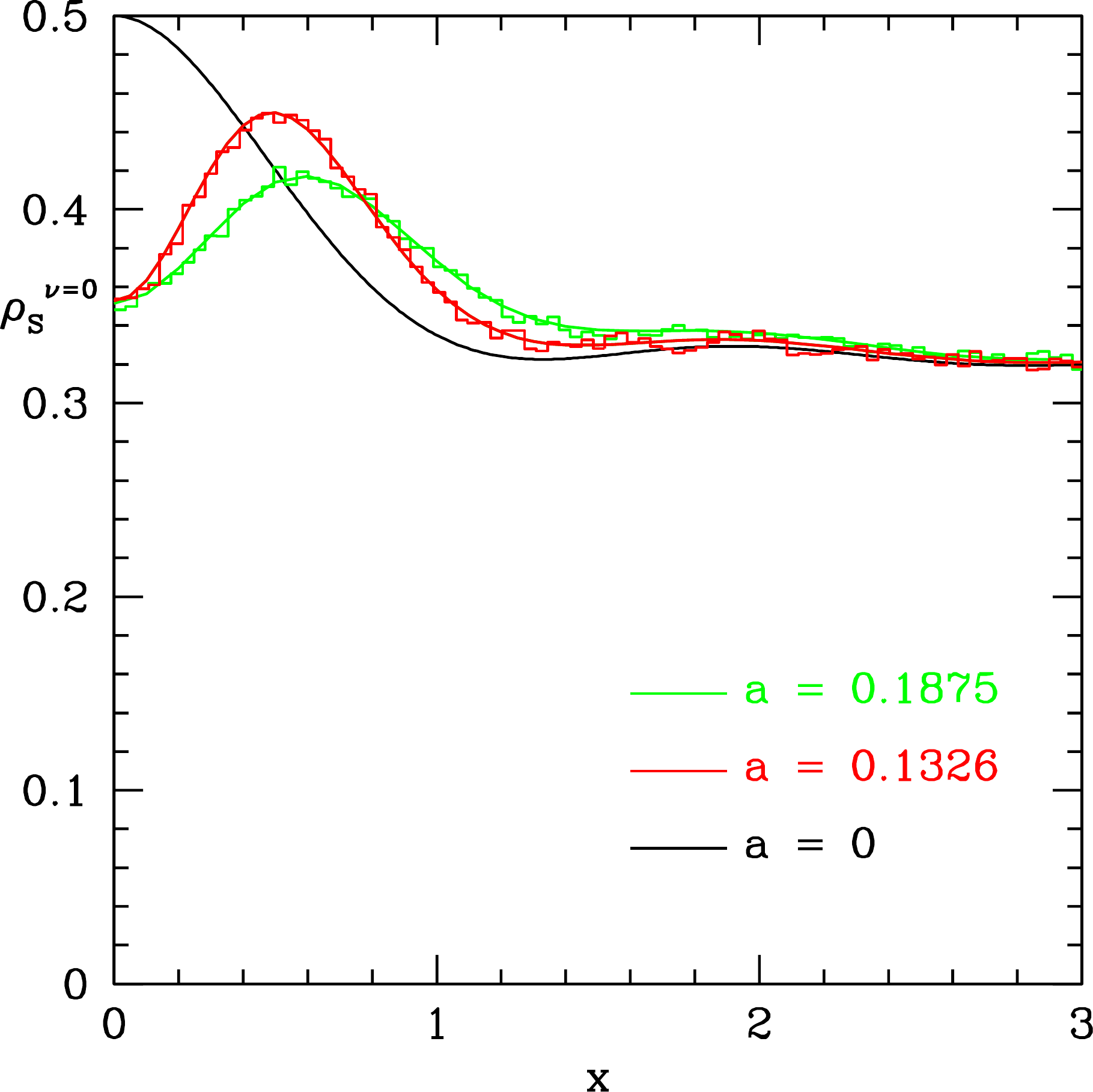}
 % \captionof{rhonu0.pdf}{A figure}
%  \caption{A figure}
  %\label{rhonu0.pdf}
\end{minipage}%
\hspace*{-1.5cm}
\begin{minipage}{.5\textwidth}
  \centering
  \includegraphics[width=.7\linewidth]{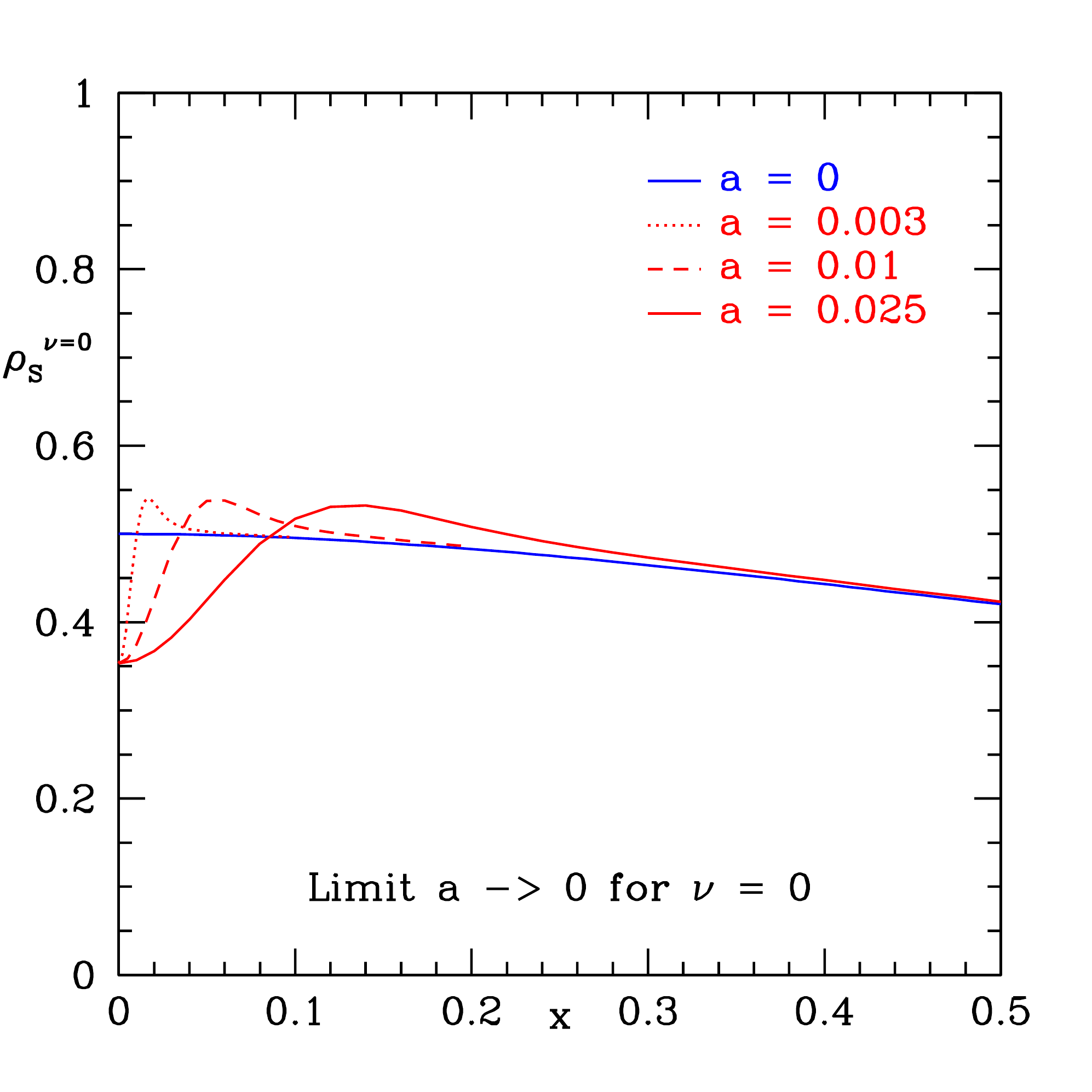}
  %\caption{Another figure}
%  \label{rhoat01.pdf}
\end{minipage}
\caption{The spectral density of $D_5$  for $z=0, m=0$ and values of $a$ as in the legend of the figures. We see that the spectral density at zero virtuality jumps by a factor $1/\sqrt{2}$ when $a\neq 0$. The solid curves denote the analytical results while the histograms represent the data of a simulation of $10^5$  $200\times 200$ random matrices. On the RHS we see the discontinuous behavior of  $\rho_5$ for very small values of $a$.}
\end{figure}
%%%%%%%

\begin{figure}
\label{bounds}
\centering
\begin{minipage}{.5\textwidth}
  \centering
  \includegraphics[width=.65\linewidth]{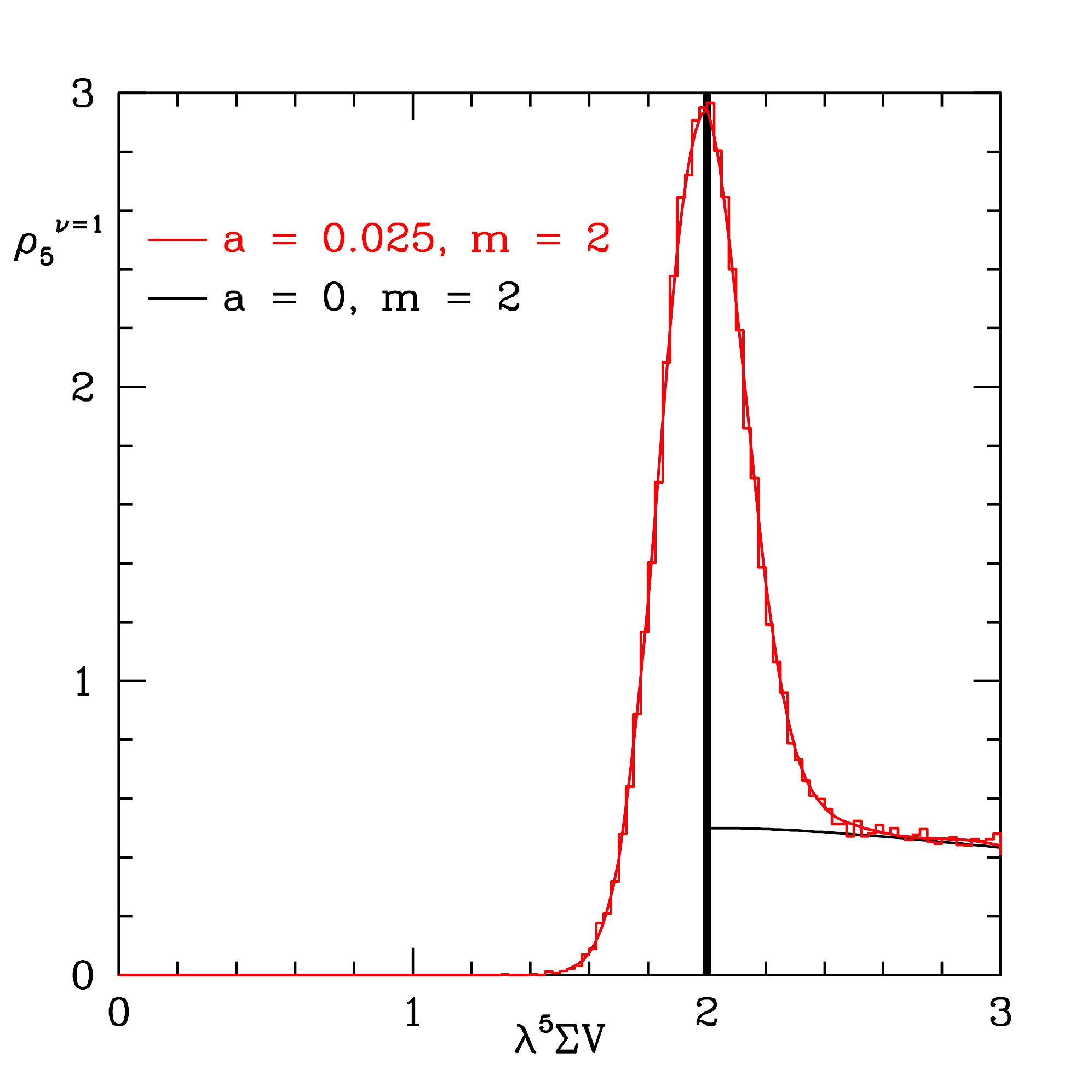}
 % \captionof{rhonu0.pdf}{A figure}
%  \caption{A figure}
  %\label{rhonu0.pdf}
\end{minipage}%
\hspace*{-1.5cm}
\begin{minipage}{.5\textwidth}
  \centering
  \includegraphics[width=.7\linewidth]{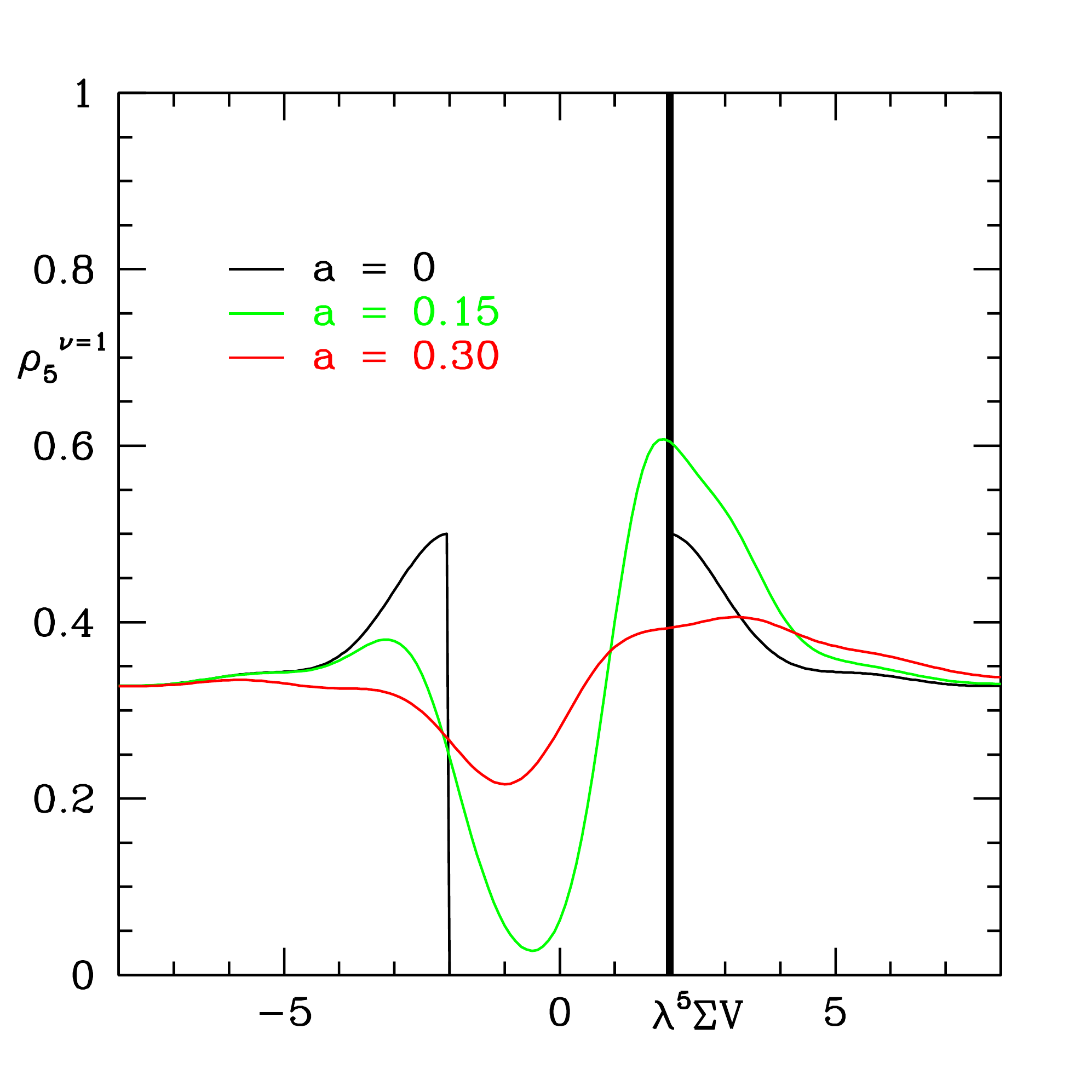}
  %\caption{Another figure}
%  \label{rhoat01.pdf}
\end{minipage}
\caption{In the left figure we show the spectral density 
of $D_5$ for $\nu =1$, $m=2$ and $a = 0.025$, and in the right
figure we display the spectral density 
of $D_5$ for $\nu =1$, $m=2$ and $a = 0$ (black), $a=0.15$ (green) 
and $a=0.3$ (red).
The numerical results 
(histogram) are compared to the analytical result (curve). Also shown
are the results for $a=0$ (black curve) with a $\delta$-peak at   
$\lambda_5 \Sigma V=2$. The $\delta$ peak corresponding to the zero modes gets broadened with a width proportional to the lattice spacing $a$.  It is interesting to see that for larger values than $a=0.3$ the peak of the zero modes has almost completely dissolved.}
\end{figure}

 %%%%%

\begin{figure}
\centering
\begin{minipage}{.5\textwidth}
  \centering
  \includegraphics[width=.65\linewidth]{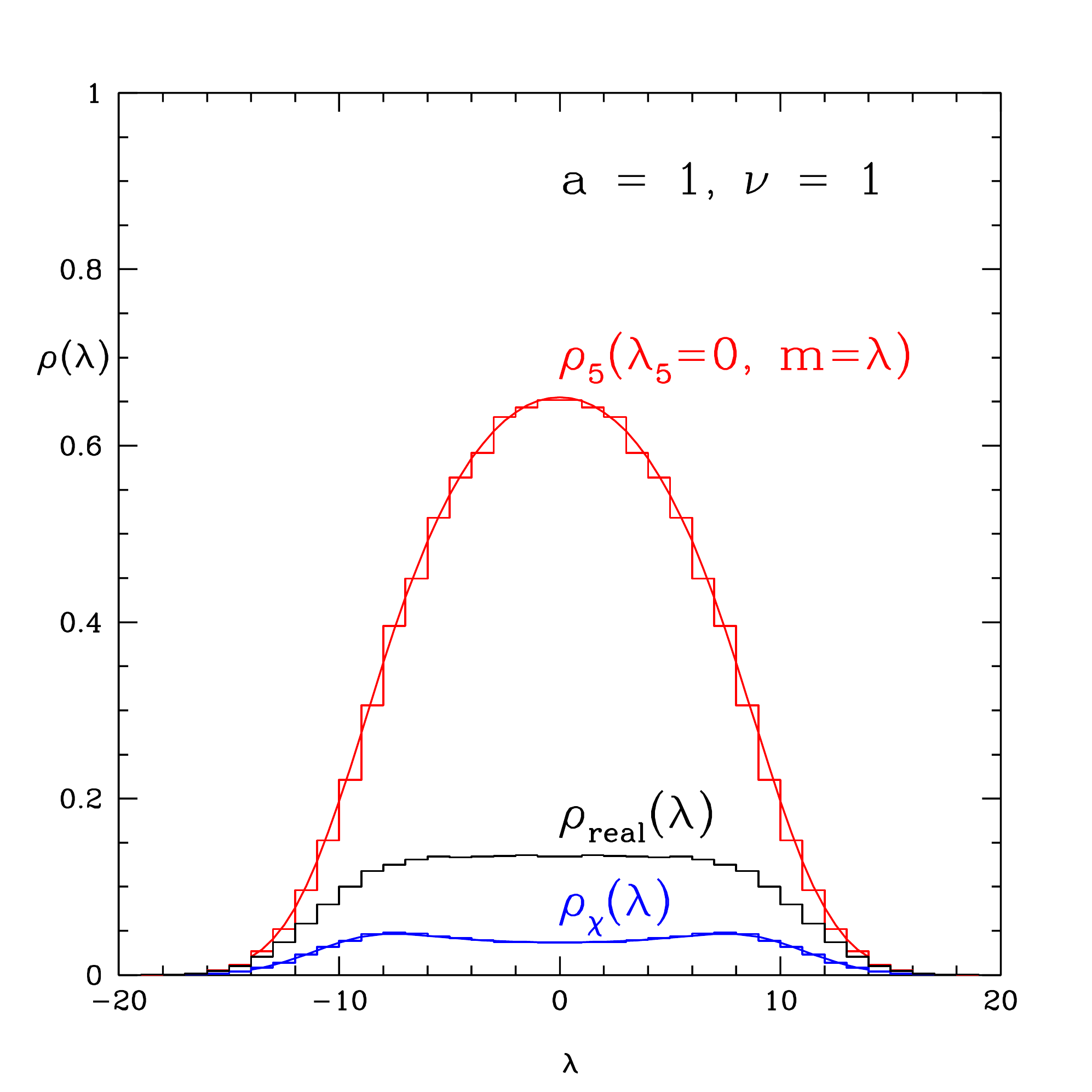}
 % \captionof{rhonu0.pdf}{A figure}
%  \caption{A figure}
  %\label{rhonu0.pdf}
\end{minipage}%
\hspace*{-1.5cm}
\begin{minipage}{.5\textwidth}
  \centering
  \includegraphics[width=.7\linewidth]{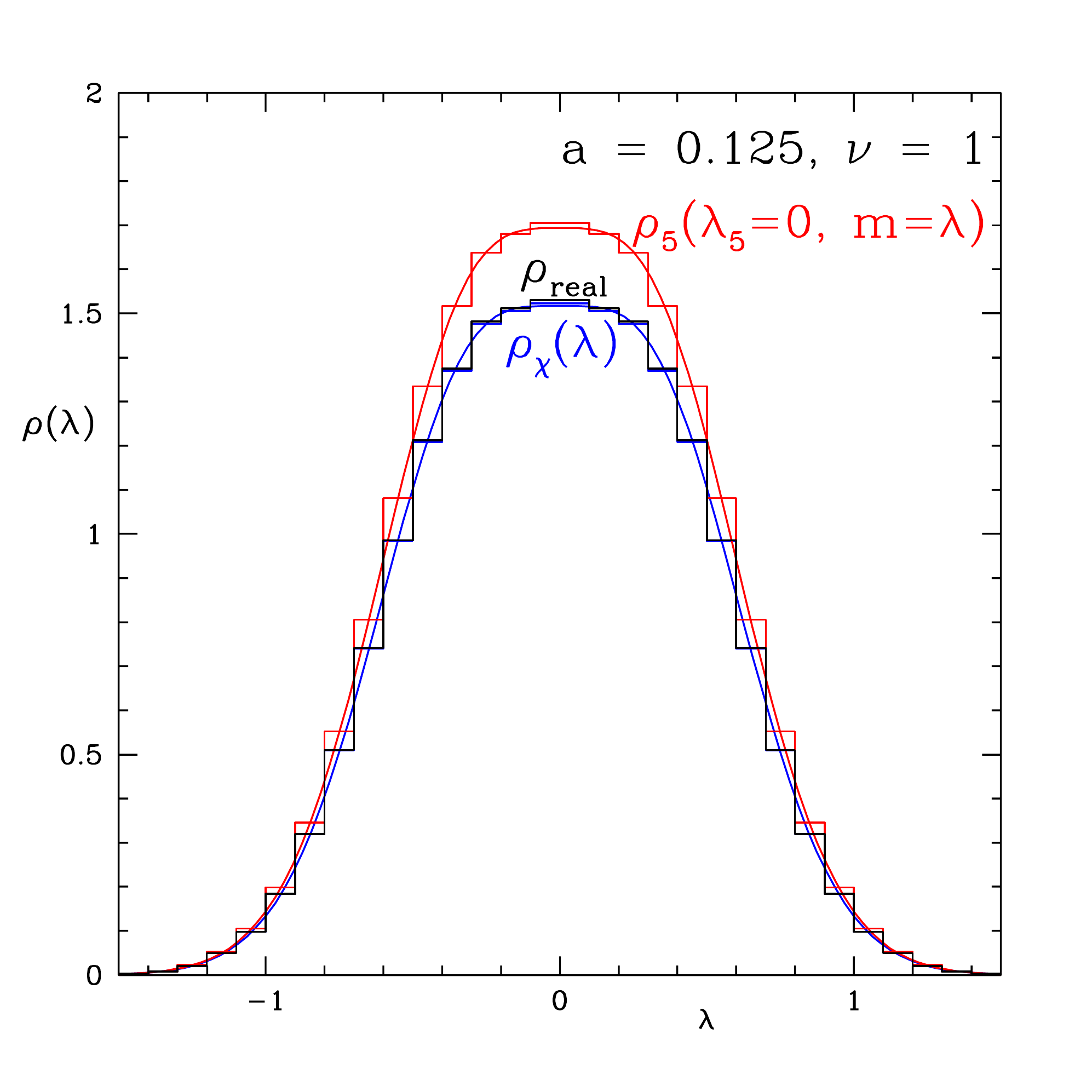}
  %\caption{Another figure}
%  \label{rhoat01.pdf}
\end{minipage}
\caption{The distribution of the chiralities (blue curves) and the
inverse chiralities (red curves) over the spectrum of $D_5$ compared
to Monte Carlo simulations for $\nu = 1$, $a=1$(left) and $ a =0.125$(right). The density $\rho_{\rm real}$ of the real eigenvalues of $D_W$ (black histogram) is always bounded by $\rho_{\chi}$ and $\rho_5(\lambda_5=0)$ and can be accurately approximated by $\rho_{\chi}$ for small values of the lattice spacing. The analytical results are in agreement with Monte-Carlo simulations (blue and red histograms).
}
\end{figure} 

 \section{Conclusions}
We have studied the microscopic spectrum of the Wilson 
Dirac operator for QCD with two colors and quarks in the fundamental 
representation. The discretization effects for $SU(2)$ QCD are mostly similar
to the case of $SU(3)$ QCD. As we have seen from the figures, the main effects are the broadening of the 
topological peak and the tail states that enter the gap 
of the Hermitian Dirac operator. Also in this case the gap closes when
entering the Aoki phase. 
Similarly to the case of ordinary QCD additional real eigenvalues could potentially be an issue in the context of $SU(2)$ QCD. These originate from the collision of complex conjugate pairs which eventually enter the real axis, and on average, we have $a^{\nu+1}$ additional real eigenvalues when we are restricted sufficiently close to the continuum limit. However, note that the suppression of these modes (for $\nu\neq 0$) is not as strong as for ordinary QCD \cite{KVZPRL}.
% In the thermodynamic limit the behavior is different, additional real eigenvalues are proportional to the lattice spacing $a$ and therefore independent of $\nu$. Meanwhile in this limit the real eigenvalue density as well as the distribution of the chiralities over the real modes have both a support of order $a^2$. 
Another noteworthy property also found for QCD with three colors is that based on the Hermiticity properties of the Dirac operator one can pinpoint \cite{KVZ-SU2} the sign of $W_8$.
In our conventions when taking into account only $W_8$ in our analysis we conclude that it should have a positive sign. The sign of $W_8$ is important because it controls if the theory enters or not the Aoki phase.
Our results provide an analytical handle on the smallest eigenvalues of the Wilson Dirac operator which could potentially seriously compromise the Monte Carlo simulation if they become almost zero. One needs to point out that the absence of level repulsion from the origin for the case of two-color QCD makes this effect much more severe compared to $SU(3)$ QCD.
\vspace*{-0.15cm}
\section*{Acknowledgments}
\vspace*{-0.35cm}
The authors wish to thank Gernot Akemann and Kim Splittorff for helpful comments. MK and SZ are financially supported by the Alexander-von-Humboldt Foundation. JV and SZ are supported by
U.S. DOE Grant No. DE-FG-88ER40388 and MK is also partially financially supported by the CRC 701: \textit{Spectral Structures and Topological Methods in Mathematics} of the Deutsche Forschungsgemeinschaft.
SZ is particularly grateful to the organizers of Moriond QCD 2015 for their financial support as well as the opportunity to present the above results in this exciting and stimulating conference.


\vspace*{-0.35cm}
\section*{References}
\vspace*{-0.35cm}
\begin{thebibliography}{99}
%\cite{Kogut:2001na}
\bibitem{Kogut} 
  J.~B.~Kogut, D.~K.~Sinclair, S.~J.~Hands and S.~E.~Morrison,
  %``Two color QCD at nonzero quark number density,''
  Phys.\ Rev.\ D {\bf 64}, 094505 (2001)
  [hep-lat/0105026].
  %%CITATION = HEP-LAT/0105026;%%
  %103 citations counted in INSPIRE as of 31 Mar 2015


%\cite{Hands}
\bibitem{Hands} 
  S.~Cotter, P.~Giudice, S.~Hands and J.~I.~Skullerud,
  %``Towards the phase diagram of dense two-color matter,''
  Phys.\ Rev.\ D {\bf 87}, no. 3, 034507 (2013)
  [arXiv:1210.4496 [hep-lat]].
  %%CITATION = ARXIV:1210.4496;%%
  %22 citations counted in INSPIRE as of 31 Mar 2015


\bibitem{kari} 
  K.~Rummukainen,
  %``QCD-like technicolor on the lattice,''
  AIP Conf.\ Proc.\  {\bf 1343}, 51 (2011)
  [arXiv: 1101.5875 [hep-lat]].
  %%CITATION = ARXIV:1101.5875;%%
  %8 citations counted in INSPIRE as of 25 Apr 2014  

\bibitem{DSV10} 
  P.~H.~Damgaard, K.~Splittorff, and J.~J.~M.~Verbaarschot,
  %``Microscopic Spectrum of the Wilson Dirac Operator,''
  Phys.\ Rev.\ Lett.\  {\bf 105}, 162002 (2010)
  [arXiv: 1001.2937 [hep-th]].
  %%CITATION = ARXIV:1001.2937;%%

%\cite{Akemann:2010em}
\bibitem{ADSV} 
  G.~Akemann, P.~H.~Damgaard, K.~Splittorff and J.~J.~M.~Verbaarschot,
  %``Spectrum of the Wilson Dirac Operator at Finite Lattice Spacings,''
  Phys.\ Rev.\ D {\bf 83}, 085014 (2011)
  [arXiv:1012.0752 [hep-lat]].
  %%CITATION = ARXIV:1012.0752;%%
  %47 citations counted in INSPIRE as of 31 Mar 2015


%\cite{KVZPRL}
\bibitem{KVZPRL} 
  M.~Kieburg, J.~J.~M.~Verbaarschot and S.~Zafeiropoulos,
  %``Eigenvalue Density of the non-Hermitian Wilson Dirac Operator,''
  Phys.\ Rev.\ Lett.\  {\bf 108}, 022001 (2012)
  [arXiv:1109.0656 [hep-lat]].
  %%CITATION = ARXIV:1109.0656;%%
  %19 citations counted in INSPIRE as of 31 Mar 2015

%\cite{lat12}
\bibitem{lat12} 
  M.~Kieburg, J.~J.~M.~Verbaarschot and S.~Zafeiropoulos,
  %``Random Matrix Models for the Hermitian Wilson-Dirac operator of QCD-like theories,''
  PoS LATTICE {\bf 2012}, 209 (2012)
  [arXiv:1303.3242 [hep-lat]].
  %%CITATION = ARXIV:1303.3242;%%
  %2 citations counted in INSPIRE as of 31 Mar 2015

%\cite{KVZlong}
\bibitem{KVZlong} 
  M.~Kieburg, J.~J.~M.~Verbaarschot and S.~Zafeiropoulos,
  %``Spectral Properties of the Wilson Dirac Operator and random matrix theory,''
  Phys.\ Rev.\ D {\bf 88}, 094502 (2013)
  [arXiv:1307.7251 [hep-lat]].
  %%CITATION = ARXIV:1307.7251;%%
  %4 citations counted in INSPIRE as of 31 Mar 2015

%\cite{KVZ-SU2}
\bibitem{KVZ-SU2} 
%\cite{Kieburg:2015vqa}
%\bibitem{Kieburg:2015vqa} 
  M.~Kieburg, J.~J.~M.~Verbaarschot and S.~Zafeiropoulos,
  %``Dirac Spectrum of the Wilson Dirac Operator for QCD with Two Colors,''
  arXiv:1505.01784 [hep-lat].
  %%CITATION = ARXIV:1505.01784;%%

\bibitem{SS}
  S.~R.~Sharpe and R.~L.~Singleton,
  %``Spontaneous flavor and parity breaking with Wilson fermions,''
  Phys.\ Rev.\  D {\bf 58}, 074501 (1998)
  [arXiv: hep-lat/9804028].
  %%CITATION = PHRVA,D58,074501;%%
  
%\cite{Verbaarschot:1994ia}
\bibitem{JV-su2} 
  J.~J.~M.~Verbaarschot,
  %``The Spectrum of the Dirac operator near zero virtuality for N(c) = 2 and chiral random matrix theory,''
  Nucl.\ Phys.\ B {\bf 426}, 559 (1994)
  [hep-th/9401092].
  %%CITATION = HEP-TH/9401092;%%
  %82 citations counted in INSPIRE as of 31 Mar 2015
\end{thebibliography}
\end{document}